\documentclass[aps,pra,preprint,10pt]{revtex4-1}  

\usepackage{amsmath}  
\usepackage{amsfonts} 
\usepackage{graphicx} 
\usepackage{color}

\newcommand{\be}[1]{\begin{equation}\label{#1}}
\newcommand{\ee}{\end{equation}}
\newcommand{\bea}[1]{\begin{eqnarray}\label{#1}}
\newcommand{\eea}{\end{eqnarray}}
\newcommand{\no}{\nonumber \\}
\newcommand{\Fig}[1]{Fig.(\ref{#1})}

\newcommand{\Eq}[1]{Eq.(\ref{#1})}

\newcommand{\Sec}[1]{Section~\ref{#1}}
\newcommand{\bsub}{\begin{subequations}}
\newcommand{\esub}{\end{subequations}}
\newcommand{\bwt}{\begin{widetext}}
\newcommand{\ewt}{\end{widetext}}

\usepackage{color}

\def\trm#1{\textrm{#1}}
\def\tit#1{\textit{#1}}

\newcommand{\om}{\omega}

\def\a0{{\alpha_0}}

\def\da0{{\dot{\alpha}_0}}

\def\myoverDefn#1#2{\hbox{\space \raise-2mm\hbox{$\textstyle{#1} \atop \scriptstyle{#2}$} }}
\def\defn{\overset{\textrm{def}}{=}}

\def\om{{\omega}}

\def\a{{\alpha}}

\def\dag{\dagger}

\def\rp{r_{P}}
\def\rp2{r_{p}^{2}}
\def\Tr{\textrm{Tr}}

\def\A{\mathcal{A}}
\newcommand{\half}{\frac{1}{2}}

\newcommand{\ket}[1]{|#1\rangle}
\newcommand{\bra}[1]{\langle #1|}

\newcommand{\IP}[2]{\langle {#1} | {#2} \rangle}
\newcommand{\EV}[2]{\langle {#1} \rangle_{#2}}

\begin{document}

\title{An examination of the extended Hong-Ou-Mandel effect \\ and considerations for experimental detection}
\author{Paul M. Alsing}\email{corresponding author: palsing@albany.edu, alsingpm@gmail.com}
\affiliation{University at Albany-SUNY, Albany, NY 12222, USA}
\author{Richard J. Birrittella}
\affiliation{Booz Allen Hamilton, 8283 Greensboro Drive, McLean, VA 22102, USA}
\affiliation{Air Force Research Laboratory, Information Directorate, 525 Brooks Rd, Rome, NY, 13411, USA}




\date{\today}

\begin{abstract}
In recent works \cite{eHOM_PRA:2022,HOMisReallyOdd:2024}  we have explored a multi-photon extension  of the celebrated two-photon Hong-Ou-Mandel (HOM) effect \cite{HOM:1987} in which the quantum amplitudes for a two-photon input to a lossless, balanced 50:50 beamsplitter (BS) undergoes complete destructive interference. In the extended Hong-Ou-Mandel (eHOM) effect \cite{eHOM_PRA:2022}
the multi-photon scattering of photons from the two input ports to the two output ports of the BS for Fock number basis input states (FS)  $\ket{n,m}_{12}$ exhibit complete destructive interference  pairwise within the quantum amplitudes containing many scattering components \cite{HOMisReallyOdd:2024}, generalizing the two-photon HOM effect. 
This has profound implications for arbitrary bipartite photonic input states  constructed from such basis states: if the input state to one input port of the BS is of odd parity, i.e. constructed from only of odd numbers of photons, then regardless of the input state to the second 50:50 BS port, there will be a central nodal line (CNL) of zeros in the joint output probability distribution along the main diagonal for coincidence detection. 
The first goal of this present work is to show diagrammatically how the extended HOM effect can be seen as a succession of multi-photon HOM effects when the latter is viewed as a pairwise cancellation of mirror image scattering amplitudes.
The second goal of this work is to explore considerations for the experimental realization of the 
extended Hong-Ou-Mandel effect. We examine the case of a single photon interfering  with a coherent state (an idealized laser) on a balanced 50:50 beamsplitter and consider prospects for experimental detection of the output destructive interference  by including additional effects such as imperfect detection efficiency, spatio-temporal mode functions, and time delay between the detected output photons. 
\end{abstract}

\maketitle 
\section{Introduction}\label{sec:Intro} 
The canonical example of quantum interference in quantum optics is the celebrated Hong-Ou-Mandel (HOM) effect \cite{HOM:1987} which is a two-photon (destructive) interference effect wherein single photons 
in either of the output beams 
of a lossless 50:50 beam splitter emerge together (probabilistically). 
(For an extensive historical review of HOM effect and its applications, see the recent review article by Bouchard \tit{et al.} \cite{Bouchard:2021}).
Detectors placed at each of the output ports will yield no simultaneous coincident clicks.
That is, the input state $\ket{1,1}_{12}$ (on modes 1 and 2),   results in the output state  $\tfrac{1}{\sqrt{2}}\left( \ket{2,0}_{12} + \ket{0,2}_{12}\right)$. The absence of the  $\ket{1,1}_{12}$ in the output is due to the complete destructive interference between the component quantum amplitudes of the  two processes (both photons transmitted, or both reflected) that potentially would lead to the state  being in the output.
The essence of this effect from an experimental point of view is that the joint probability $P_{12}(1,1)$  for detecting one photon in each of output beams vanishes, i.e.   $P_{12}(1,1)=0$.
As is well appreciated now, the HOM effect occurs because the quantum amplitude for both input photons to transmit through the BS has equal magnitude, but opposite sign, to that of  the quantum amplitude for both input photons to reflect off the beamsplitter. This cancelation of quantum amplitudes was stressed by  Glauber \cite{Glauber:1995}, who pointed out that  in spite of its name, ``multiphoton interference" does not involve the interference of photons. Rather, it is  the addition of the quantum amplitudes (themselves being complex numbers, acting effectively as the square roots of a probabilities with complex phases) associated with these states that give rise to interference effects. 
The process that gives rise to such two-mode states of light via beam splitting is known 
as multiphoton interference \cite{Ou:1996,Ou:2007,Ou_Book:2017},  
and serves as a critical element in several applications including quantum optical interferometry \cite{Pan:2012}, and quantum state engineering where beam splitters and conditional measurements are utilized to perform post-selection techniques such as photon subtraction \cite{Dakna:1997,Carranza:2012,Magana-Loaiza:2019}, photon addition  \cite{Dakna:1998}, and photon catalysis \cite{Lvovsky:2002, Bartley:2012, Birrittella:2018}.

\begin{figure*}[th]
\begin{center}
\begin{tabular}{cccc}
\hspace{-0.65in}
\includegraphics[width=1.85in,height=1.35in]{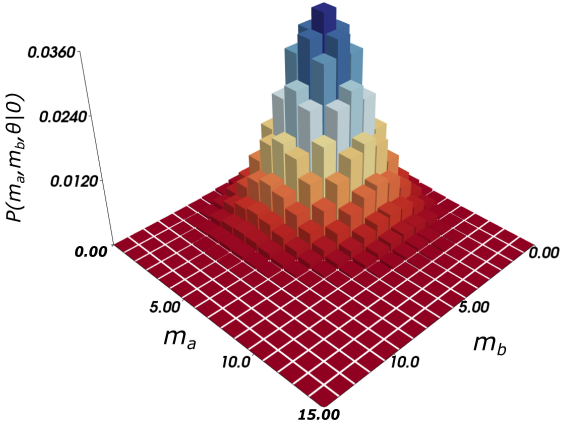}       &  
\includegraphics[width=1.85in,height=1.35in]{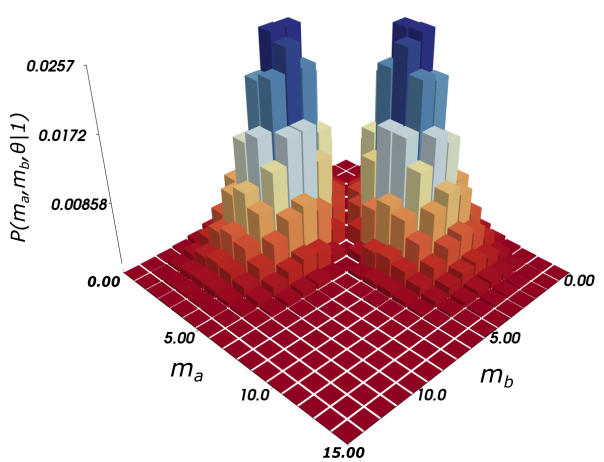}   & 
\includegraphics[width=1.85in,height=1.35in]{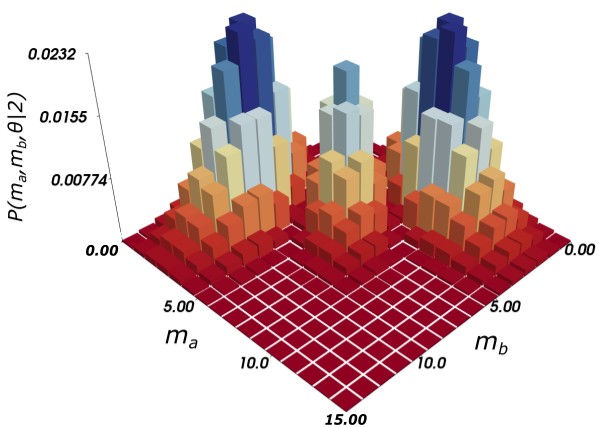} & 
\includegraphics[width=1.85in,height=1.35in]{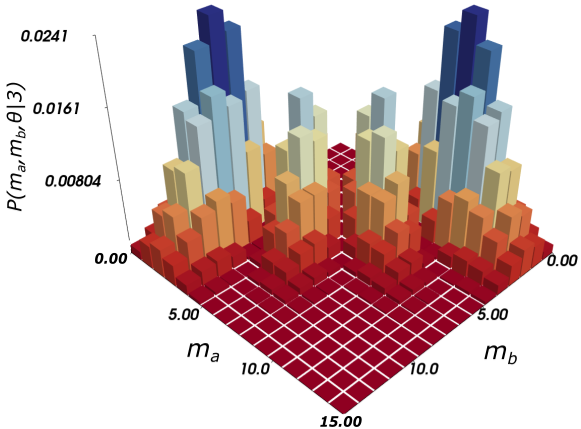} \\
\hspace{-0.65in}
\includegraphics[width=1.85in,height=1.35in]{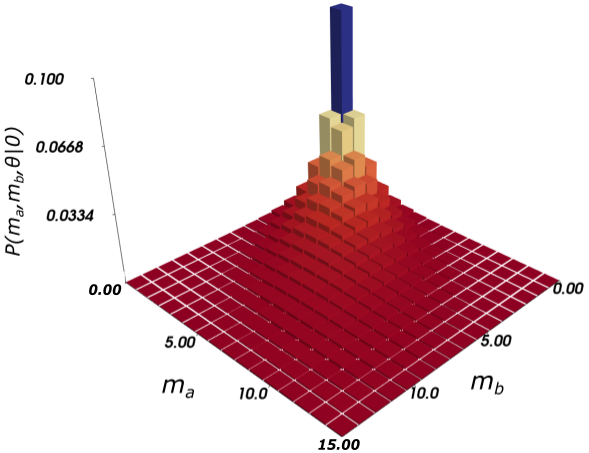}       &  
\includegraphics[width=1.85in,height=1.35in]{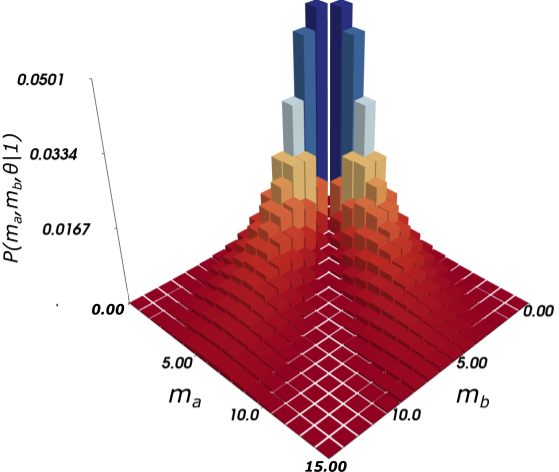}   & 
\includegraphics[width=1.85in,height=1.35in]{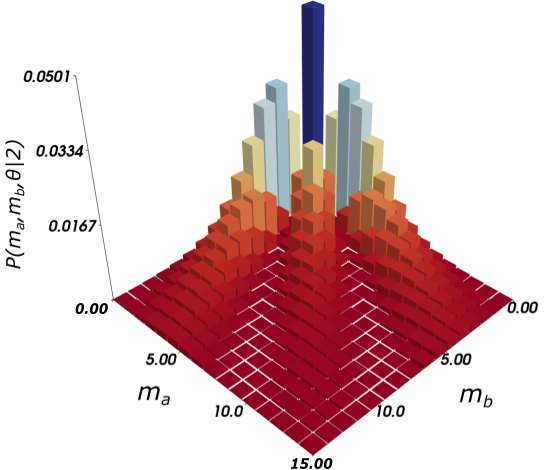} & 
\includegraphics[width=1.85in,height=1.35in]{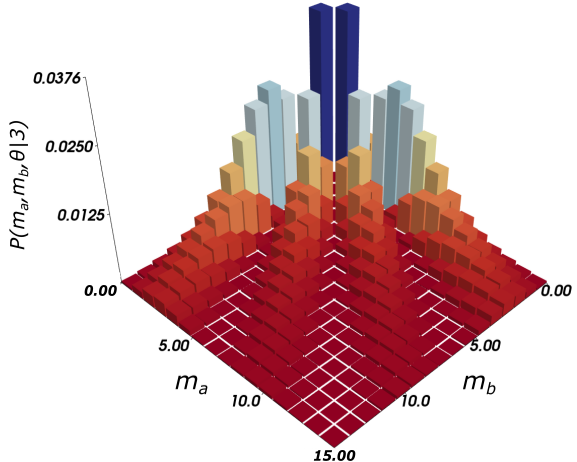}   
\end{tabular}
\end{center} 
\caption{Joint output probability 
$P(m_a, m_b|n)$ to measure $m_a$ photons in mode-1 and $m_b$ photons in mode-2
from a 50:50 BS
for input  Fock number states (FS) $\ket{n}_1$ in mode-1, for $n=\{0,1,2,3\}$ (top row, left to right),
 and  an input coherent state (CS) $\ket{\beta}_2$ in mode-2, with mean number of photons 
with  $\bar{n}_2 =9$.  A central nodal line (CNL) of zeros for inputs $\ket{n,\beta}_{12}$ 
is observed for odd $n=\{1,3\}$ indicating destructive interference of coincidence 
detection on all output FS/FS $\ket{m',m'}_{12}$.
No CNL is observed for input states with even $n=\{0,2\}$ , indicating non-zero coincidence detection.
(bottom row) Same as top row, but now with the CS mode-2 input state replaced by a mixed thermal state $\rho^{\trm{thermal}}_2$ of average photon number $\bar{n}_2 =9$.
}
\label{fig:eHOM:fig2:fig3:left:column:1photon:CS}    
\end{figure*}

Recently, the authors in \cite{eHOM_PRA:2022} have shown a multi-photon extension of the HOM effect which they termed the extended HOM effect (eHOM), in which complete destructive interference of the quantum amplitudes for coincident detection output states 
$\ket{\tfrac{n+m}{2}, \tfrac{n+m}{2}}_{12}$
for a balanced (50:50) lossless BS
occurs for any Fock (number) state  input $\ket{n,m}_{12}$ (FS) when  $n$ and $m$ are both \tit{odd}, but does not occur when $n$ and $m$ are both \tit{even} (clearly their is no possibility for coincidence detection if $(n,m)$ are either (even,odd) or (odd,even)). Since the  input states $\ket{n,m}_{12}$ form a complete orthonormal dual basis for all bipartite states (both pure and mixed), the eHOM effect implies that for any odd parity state (comprised of only an odd number of photons) entering the BS input port 1, then regardless of the state entering input port 2, either pure or mixed, there will be a line of zeros, a central nodal line (CNL), down the diagonal of the joint output probability for coincidence detection. 
This is illustrated in  \Fig{fig:eHOM:fig2:fig3:left:column:1photon:CS}  showing the joint output probability $P(m_a, m_b|n)$ to detect $m_a$ photons in output port 1 and $m_b$ photons in output port 2, given that a FS $\ket{n}_1$ enters input port 1, for $n=0,1,2,3$ photons. In the top row, the input into port 2 is a coherent state (CS), e.g. an idealized laser, of mean number of photons $\bar{n}_2 = |\beta|^2=9$. The CNL in the second and fourth figures (from left to right) is clearly visible \cite{eHOM:PNCs:note}. 
The bottom row of  \Fig{fig:eHOM:fig2:fig3:left:column:1photon:CS} is similar to the top row, except now the input CS state is replaced by a mixed thermal state $\rho^{\trm{thermal}}$, again of mean photon number 
$\bar{n}_2 =9$, illustrating the universality of the eHOM effect.  

The case of the input of a single photon and a CS $\ket{1,\beta}_{12}$ on a BS was discussed over the years by many authors, most notably Ou (of HOM fame) in 1996 \cite{Ou:1996} (and in subsequent books) \cite{Ou:2007,Ou_Book:2017}, and by Birrittella Mimih and Gerry \cite{BMG:2012} in 2012, but never fully explored as in \cite{eHOM_PRA:2022} (which also examined  non-balanced 
lossless BS configurations as well).  
In his book ``Quantum Optics for Experimentalists," \cite{Ou_Book:2017}, Ou coins the term "Generalized HOM effect," (Chapter 8.3.2) in which judicious choices of the transmission coefficient (for a non-balanced lossless BS) can lead to destructive interference on chosen non-diagonal coincidence states (vs the balanced lossless BS and universality of the CNLs discussed in this work).
%
The early work of 
Lai, Bu\u{z}ek and Knight (LBK) \cite{Lai:1991}  looked at the BS transformation on dual FS inputs to a fiber-coupler BS, including scattering losses (due to sidewall roughness).
%
The work that comes closest to nearly addressing the eHOM effect was that by Campos, Saleh and Teich \cite{Campos:1989} in their extensive study of the $SU(2)$ properties of a lossless beam splitter. Once again, conditions were discussed for obtaining isolated zeros in the joint output probability distribution, for interesting special cases, but not in all generality.


In this work, we wish to consider the prospects for the experimental realization of the eHOM effect for the case of FS/CS input $\ket{1,\beta}_{12}$ to a balanced 50:50 BS. 
So far in the prior works of the authors  \cite{eHOM_PRA:2022,HOMisReallyOdd:2024} the eHOM effect has been discussed in an idealized form, which has implicitly assumed that (i) the photons from both input ports arrive at the BS simultaneously, (ii) the detectors for measuring the output photons have unit efficiency, and the (iii) the photons involved are treated as monochromatic plane waves. Here we will consider the implications of these common detrimental effects  likely encountered in any attempt at an experimental realization of the eHOM effect, and their degradation on the resulting destructive interference.

The outline of this paper is as follows;
%
In \Sec{sec:Review:HOMisOdd} we briefly review the results of \cite{HOMisReallyOdd:2024} to show
how the eHOM effect for FS/FS inputs $\ket{n,m}_{12}$ can be interpreted, solely diagrammatically, as a multi-photon generalization of the two-photon HOM effect. This is accomplished by keeping track of how input photons are scattered (transmitted or reflected) into output modes, and the $(-1)$ signs they encounter. 
%
In \Sec{sec:1:beta:input} we begin our  examination of the  prospects for an experimental realization of the eHOM effect on the $\ket{1,1}_{12}$ output coincident state, for an input  consisting of  a single photon FS in mode-1 and a CS (idealized laser)  in mode-2. In this section we include the effects of finite (imperfect) detection efficiency.
In \Sec{sec:Rempe:calc} we include the effects of the wavepacket nature of FS input photons by including
 spatio-temporal modes functions in our experimental examination, as well as the effect of a possible time delay between the detected output photons. For the FS/CS input, we first consider these effects on the detection of the output coincidence state  $\ket{1,1}_{12}$, before examining the more involved case of detection of the 
 general output diagonal state  $\ket{N,N}_{12}$, involving photon counting theory/expressions. 
 In \Sec{sec:Conclusion} we present a summary and conclusion of our results.
 
\section{A diagrammatic interpretation of the extended HOM effect}\label{sec:Review:HOMisOdd}
\begin{figure*}[th]
\begin{center}
\includegraphics[width=3.0in,height=2.25in]{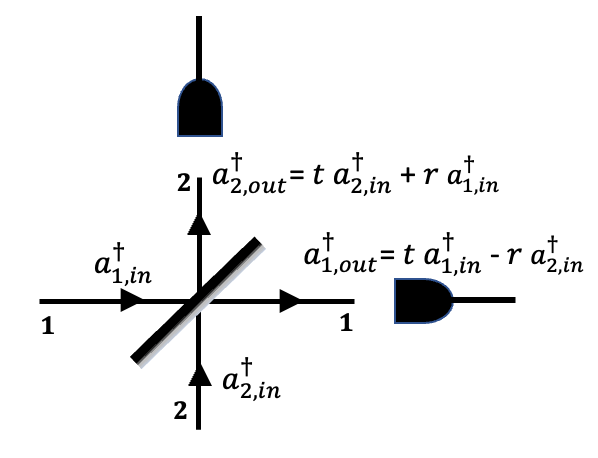}
\end{center} 
\caption{Output creation operators for mode-1 and mode-2 in terms of the input creation operators for a lossless beamsplitter (BS), with transmission coefficient $t$ and reflection coefficient $r$ such that 
$|t|^2 + |r|^2\equiv T+R=1.$ In the convention employed in this work, $t$ and $r$ are taken to be real, and the $(-1)$ sign is associated with an input mode-2 photon reflecting into the output mode-1.
%
}
\label{fig:eHOM:BS:schematic}    
\end{figure*}

Any two-port optical device such as a lossless BS is described  
by the unitary transformation $U$ of the input mode creation  operators $a^\dag_{1,in}, a^\dag_{2,in}$ to the output modes $a^\dag_{1,out}, a^\dag_{2,out}$ given by
\be{BS:S:out:in:terms:of:in}
\hspace{-0.5in}
\vec{a}^\dag_{out}=
\left[
\begin{array}{c}
a^\dag_{1,out} \\
a^\dag_{2,out} 
\end{array}
\right] = 
U
\left[
\begin{array}{c}
a^\dag_{1,in} \\
b^\dag_{2,in}
\end{array}
\right] 
U^\dag 
=
\left[
\begin{array}{cc}
S_{11} & S_{12}\\
S_{21} & S_{22}
\end{array}
\right]
\,
\left[
\begin{array}{c}
a^\dag_{1,in} \\
a^\dag_{1,in}
\end{array}
\right] 
\equiv
S\, \vec{a}^\dag_{in}
=
\left[
\begin{array}{c}
S_{11}\,a^\dag_{1,in} +S_{12}\,a^\dag_{2,in}\\
S_{21}\,a^\dag_{1,in} +S_{22}\,a^\dag_{2,in}\end{array}
\right].
\ee
We see that $S_{ij}$ is interpreted as the scattering of an input photon from mode-$j$ 
into the output mode-$i$. (Note: in the subsequent figures we have included a ``backward circumflex arrow" from $j$ to $i$ to mnemonically remind the reader of the direction of the mode scattering in the amplitude $S_{ij}$). Unitarity of the $U$ requires the unitarity of $S$ which can be expressed as the requirement of the orthonormality of the columns (and rows) of $S$, yielding \cite{Skaar:2004}
\be{unitarity:of:S}
|S_{11}|^2 + |S_{12}|^2 =1,\quad 
|S_{21}|^2 + |S_{22}|^2 =1, \quad
S_{11}\,S^*_{12} + S_{21}\,S^*_{22}=0.
\ee
Taking the absolute value of the last equation in \Eq{unitarity:of:S} and using the first two equations yields 
$|S_{11}| =|S_{22}|$ and $|S_{12}| = |S_{21}| = \sqrt{1-|S_{11}|^2}$. Finally, writing each amplitude as 
$S_{ij} = |S_{ij}|\,e^{\theta_{ij}}$, the last equation in \Eq{unitarity:of:S} yields the phase condition
$\theta_{11}+\theta_{22}=\theta_{12}+\theta_{21}+\pi$.
While there appears in the literature and textbooks
many different phase conventions for a BS \cite{Loudon:2000, Skaar:2004,Agarwal:2013,Ou_Book:2017,Gerry_Knight:2023},
none of the results of the HOM or eHOM effect depends on a particular choice of phase, and so for convenience, we chose to use an anti-symmetric $2\times 2$ rotation matrix, with real entries  
and $S_{21}=r = -S_{12}$ \cite{eHOM_PRA:2022}.   Thus, the transformation of the mode creation operators in  \Eq{BS:S:out:in:terms:of:in} is illustrated in \Fig{fig:eHOM:BS:schematic} for this phase convention, which allows us to track a single $-1$ sign change ($S_{12}$) associated with the scattering (reflection) of an input mode-2 photon off the bottom of the BS, into mode-1. 
Again, note that in general $S_{ij}$ is defined to be \cite{Skaar:2004} the amplitude for a single photon to scatter from \tit{input} mode-$j$ to \tit{output} mode-$i$, as given by the rightmost BS transformation given in \Eq{BS:S:out:in:terms:of:in}.

\subsection{The Hong-Ou-Mandel effect}\label{sec:2:photon:input}
\begin{figure*}[th]
\begin{center}
\includegraphics[width=4.0in,height=1.75in]{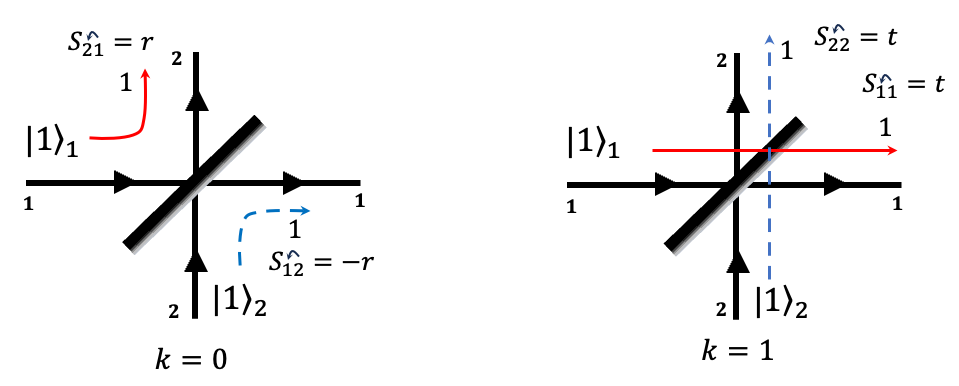}
\end{center} 
\caption{The 2-photon HOM effect with input $\ket{1,1}_{12}$ illustrating the two scattering amplitudes (left) $A_{k=0}$ both photons reflecting off the BS, and (right) $A_{k=1}$ both photons transmitting through the BS. 
Both amplitudes have equal magnitudes,  yet opposite signs, that combine to create destructive interference on the coincident output state $\ket{1,1}_{12}$. 
In general, the integer $k\in\{0,1,\ldots,n\}$ designates the number of  photons that transmit from input mode-1 to output mode-1 (with amplitude $(S_{11})^k$)
for the FS/FS input state $\ket{n,m}_{12}$.
}
\label{fig:HOM:FS:FS:1:1}    
\end{figure*}

Before examining the eHOM effect, 
let us first consider the standard textbook discussion of the HOM effect
\cite{Agarwal:2013,Ou_Book:2017,Gerry_Knight:2023} in terms of the scattering amplitudes of the two input photons.
In \Fig{fig:HOM:FS:FS:1:1}  we consider the two-photon input state $\ket{1,1}_{12}$, 
for the HOM effect \cite{HOM:1987}, and consider the 
total amplitude $A$ for coincidence on the output state $\ket{1,1}_{12}$. The total amplitude is composed of two components $A = A_{k=0}+A_{k=1}$, where, in general, $k=\{0,\ldots,n\}$ (note: without loss of generality we will only consider the cases where $n\le m$) indicates the number of photons transmitted from input mode-1 to output mode-1 (giving rise to the contribution $(S_{11})^k$ in $A_k$).  As is well known now, the HOM effect, or the complete destructive interference of the quantum amplitudes for the output on the $\ket{1,1}_{12}$, is given by the sum of the contributions where (i) both photons are reflected into opposite numbered modes by the BS, with amplitude $A_{k=0} = S_{21}\,S_{12} = (r)(-r)= -r^2$, and both photons transmitted into their originating numbered modes, with amplitude $A_{k=1} = S_{11}\,S_{22} = t^2$, as shown in  \Fig{fig:HOM:FS:FS:1:1}. 
The crucial $(-1)$ sign in the $A_{k=0}$ amplitude comes from the contribution 
$(S_{12})^1 = -r$ of the single input photon (the exponent of $S_{12}$) from mode-2  reflecting into mode-1.  
For a 50:50 BS where $t=r=\tfrac{1}{\sqrt{2}}$
these two amplitudes have equal magnitude, but opposite sign, and hence sum to zero:
$A = A_{k=0} + A_{k=1} = (-r^2 + t^2)\overset{|t|=|r|}{\to}0$ 
(complete destructive interference) \cite{note:on:choice:of:S:in:HOM:v2}. 
This is the famed HOM effect \cite{HOM:1987}. Of course, this is the ``textbook version" of the HOM effect, since we have implicitly assumed, that both photons 
(i) are completely indistinguishable (e.g. monochromatic in frequency, same spatial mode profile, etc\ldots), 
(ii) have arrived at the BS at the same time (no relative time delay), and 
(iii) have been detected at the same time (no detection time difference). All these effects can be incorporated into the analysis of the HOM effect \cite{Legero_Rempe:2003}, and we will examine some of these considerations in later sections.

A subtle feature in \Fig{fig:HOM:FS:FS:1:1}, that is obscured by the employment of only two input photons in the HOM effect, is the $k=0$ and $k=1$ component scattering amplitudes diagrams are ``mirror images" of each other. By this we mean the following. 
Focusing on mode-1, we see that the number of  photons 
reflected  from input mode-1 to output mode-2 (red solid arrow) for $k=0$ in \Fig{fig:HOM:FS:FS:1:1}(left) is equal to the number of photons transmitted  from input mode-1 to output mode-1(red solid arrow) for $k=1$ in  \Fig{fig:HOM:FS:FS:1:1}(right). Similarly for mode-2, the number of photons reflected  from input mode-2 to output mode-1 (blue dashed arrow) for $k=0$ in \Fig{fig:HOM:FS:FS:1:1}(left) is equal to  to the number of photons transmitted  from input mode-2 to output mode-2 (blue dashed arrow) for $k=1$ in  \Fig{fig:HOM:FS:FS:1:1}(right).

While this swapping of the number of photons that reflect/transmit in one diagram to number that transmit/reflect respectively in a ``mirror image" diagram, but with an overall relative minus sign, may appear trivial when only two photons are involved, it actually lies at the heart of the eHOM destructive interference effect. As discussed diagrammatically in \cite{HOMisReallyOdd:2024}, for general FS/FS inputs $\ket{n,m}_{12}$ with $n$ and $m$ both \tit{odd} (taking $n<m$ for concreteness, without loss of generality) there are an even number $n+1$ of component scattering amplitudes $A_k$, $k\in\{0,1,\ldots,n\}$ comprising the total amplitude $A = \sum_{k=0}^{n} A_k$, which \tit{cancel separately in pairs} on the coincident output state $\ket{\tfrac{n+m}{2},\tfrac{n+m}{2}}_{12}$ via
$A = \sum_{k=0} ^{(n-1)/2} (A_k + A_{n-k})$. That is, each pair of 
scattering amplitudes $A_k + A_{n-k}$ are separately  ``mirror images" of each other, with equal magnitude and opposite relative minus sign when the BS is balanced (50:50), 
and thus cancel each other $A_k + A_{n-k}=0$, for $k\in\{0,1,\ldots,\tfrac{n-1}{2}\}$. Thus, one can interpret the eHOM effect as a series of  HOM effects happening simultaneously on $\tfrac{n+1}{2}$ pairs of multi-photon scattering amplitudes. This can be diagrammatically shown, as in the next subsection.

Lastly, for $(n,m)$ both even there are an \tit{odd} number of component scattering amplitudes, 
and (i) mirror image diagrams now have the \tit{same} sign and therefore constructively interfere, 
and (ii) there is an ``unpaired" non-zero scattering amplitude that is not able to cancel with any other diagram. Both these conditions imply that there cannot be destructive interference on the output coincidence state if $(n,m)$ are both even \cite{eHOM_PRA:2022,HOMisReallyOdd:2024}.

\subsection{The extended Hong-Ou-Mandel effect}\label{sec:4:photon:input:3:5:output:4:4}
\begin{figure*}[!ht]
\begin{center}
\includegraphics[width=5.5in,height=1.75in]{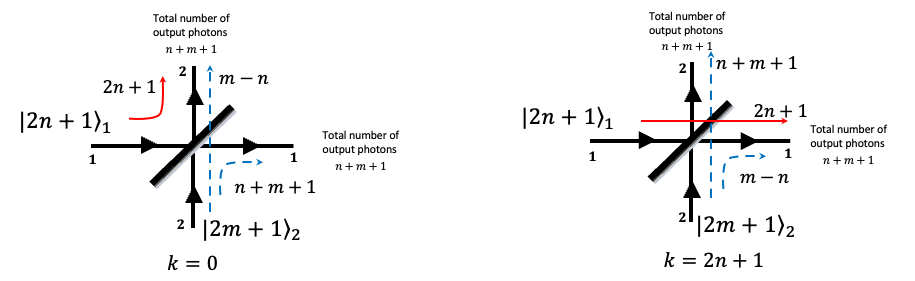}
\end{center} 
\caption{The odd-odd-photon eHOM effect with input state $\ket{2n+1,2m+1}_{12}$ illustrating the two ``outer" scattering amplitudes $A_{k=0}$ and $A_{k=2n+1}$ (for $k\in\{0,1,\ldots,2n+1\}$),
with equal magnitude and opposite signs (when $|t|=|r|$) that cancel each other, contributing to the  complete destructive interference on the coincident output state $\ket{n+m+1,n+m+1}_{12}$.
}
\label{fig:eHOM:FS:FS:odd:odd:in:0:2np1:ampls}    
\end{figure*}

\begin{figure*}[!ht]
\begin{center}
\includegraphics[width=5.5in,height=1.75in]{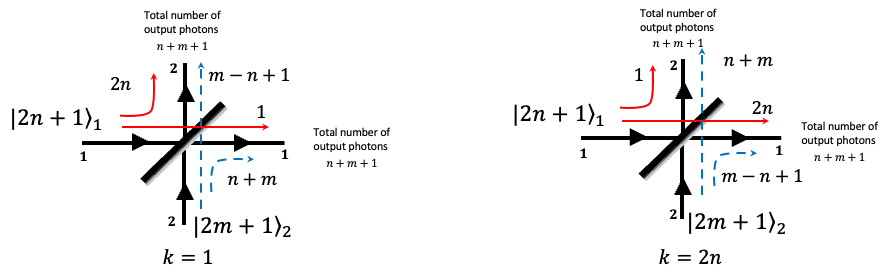}
\end{center} 
\caption{The odd-odd-photon eHOM effect with input $\ket{2n+1,2m+1}_{12}$ illustrating the two ``penultimate outer" scattering amplitudes $A_{k=1}$ and $A_{k=2n}$ (for $k\in\{0,1,\ldots,2n+1\}$), with equal magnitude and opposite signs (when $|t|=|r|$) that cancel each other, contributing to the  complete destructive interference on the coincident output state $\ket{n+m+1,n+m+1}_{12}$. 
}
\label{fig:eHOM:FS:FS:odd:odd:in:1:2n:ampls}    
\end{figure*}

To show the pairwise cancellation of mirror image scattering amplitudes 
 $(A_{k} + A_{k=n-k})\overset{t=r}{\to} 0$
in the previous subsection for the case of a FS/FS input state $\ket{n,m}_{12}$ with
$(n,m)$ both odd, consider  
%
\Fig{fig:eHOM:FS:FS:odd:odd:in:0:2np1:ampls} and
\Fig{fig:eHOM:FS:FS:odd:odd:in:1:2n:ampls}  with the general odd-odd FS/FS input 
$\ket{2n+1,2m+1}_{12}$, with output coincident state $\ket{n+m+1,n+m+1}_{12}$. 
%

\Fig{fig:eHOM:FS:FS:odd:odd:in:0:2np1:ampls} considers the ``outermost" pair of amplitude diagrams
$A_{k=0}+A_{2n+1}$ (the minimum and maximum number of photons transmitted from input mode-1 to output mode-1, respectively), 
where the symmetry dictates the equality of the combinatorial factors, $C_0=C_{2n+1}$ \cite{HOMisReallyOdd:2024}. 
The amplitude for the leftmost diagram is given by
$A_{k=0}=C_0\,(S_{11})^{0}\,(S_{21})^{2n+1}\,(S_{22})^{m-n}\,(S_{12})^{n+m+1}=
C_0\,t^{n+m+1}\,r^{n+m+1}\,(-1)^{n+m+1}$, while the amplitude for the rightmost diagrams is given by
$A_{k=2n+1} = C_{2n+1}\,(S_{11})^{2n+1}\,(S_{21})^{0}\,(S_{22})^{n+m+1}\,(S_{12})^{m-n}=
C_{2n+1}\,t^{3n+m+2}\,r^{m-n}\,(-1)^{m-n}$.
The sum of amplitudes for these two diagrams can be factored into the form
$
A_{k=0} +A_{k=2n+1} = C_0\,t^{n+m+1}\,r^{m-n}\,(-1)^{m-n}\,
$
$
[
r^{2n+1}\,(-1)^{2n+1} + t^{2n+1}
] \overset{|t|=|r|}{\to}0
$.
The crucial point of this last expression shows  that while both diagrams incur multiple powers of $(-1)$ due to the scattering of mode-2 photons into mode-1 ($S_{12}$), the \tit{relative sign} between the two diagrams (in the square brackets) is $(-1)^{2n+1}\equiv -1$. Thus, the pair of mirror image diagrams cancel each other for a 50:50 BS, with equal magnitude, but opposite sign.

\Fig{fig:eHOM:FS:FS:odd:odd:in:1:2n:ampls} considers the ``next innermost pair" of mirror image amplitude diagrams, where in a similar calculation to the above, we have 
$
A_{k=1} +A_{k=2n} = C_1\,t^{m-n+2}\,r^{m-n+1}\,(-1)^{m-n+1}\,
[
r^{4n-2}\,(-1)^{2n-1} + t^{4n-2}
] \overset{|t|=|r|}{\to}0
$.
Again, the symmetry of the mirror image diagrams dictates the equality of the combinatorial factors, $C_1=C_{2n}$. This time the \tit{relative sign} between the two diagrams is $(-1)^{2n-1}\equiv -1$ (in the square brackets). 
Thus, once again for a 50:50 BS this pair of diagrams exhibit equal magnitudes (though different in value from that of the outermost pair) and  opposite signs. 

This cancellation $A_k+A_{2n+1-k}=0$ of mirror image scattering amplitudes for a 50:50 balanced beamsplitter occurs \tit{separately} for each pair  for $k\in\{0,1,\ldots,n\}$.
From these observations, a simple, general analytical proof can be constructed straightforwardly, and appears in  \cite{HOMisReallyOdd:2024} (see also Appendix B of \cite{eHOM_PRA:2022} for a formal, but less intuitive, proof).


\section{Issues for consideration for the observation of the \lowercase{e}HOM effect with a Fock state/Coherent State input to a 50:50 BS}\label{sec:1:beta:input}
We now turn our attention to a consideration of experimentally realizing the eHOM effect in a laboratory setting.
In a realistic experiment one has to contend with the prospects of imperfect detector efficiency, the influence of the photon mode functions in wavepackets, and potential time delay between photon detection in the output ports of the BS. In this section we consider two of these aspects in the context of a possible experimental realization of the eHOM effect for the case of a FS/CS input $\ket{1,\beta}_{12}$ with a single  photon $\ket{1}_1$ entering port-1 of a 50:50 BS, and a coherent state  
$\ket{\beta}_2  = e^{-|\beta|^2/2} \,\sum_{m=0}^\infty \tfrac{(\beta)^m}{\sqrt{m!}}\,\ket{m}_2$ of (complex) amplitude $\beta$, with mean photon number $\bar{n}_2 = |\beta|^2\,$
\cite{Scully_Zubairy:1997,Loudon:2000,Agarwal:2013,Ou_Book:2017,Gerry_Knight:2023}.  
The coherent state represents an idealized, continuous wave (CW) laser of fixed frequency $\omega_2$, and has the property that it is an eigenstate of the mode-2 annihilation operator, $a_2\,\ket{\beta}_2 = \beta\,\ket{\beta}_2$. 
The single photon FS could be generated for example by the heralding on one component ($1'$) of a weak two-mode squeezed state
$\ket{TMSS}_{11'} = \frac{1}{\cosh(r)}\sum_{n=0}^\infty \tanh^n(r)\ket{n,n}_{11'}$ where $r$ is the squeezing parameter. This state can be generated for example, by the nonlinear processes of spontaneous parametric down-conversion or by four-wave mixing \cite{Boyd:1991,Scully_Zubairy:1997,Agarwal:2013,Ou_Book:2017,Gerry_Knight:2023,Boyd:1991}.
For a weak field TMSS, a detection in mode-$1'$ ``heralds" the correlated presence of a single photon in mode-1, which can then be fed forward into the mode-1 input port of a 50:50 BS.

\subsection{Imperfect detection efficiency}\label{subsec:Imperfect:Detectors}
The probability $P(N_1,N_2)$ to perfectly detect $N_1$ photons from mode-1 and $N_2$ photons from mode-2 from the output of a lossless 50:50 BS is given by $P(N_1,N_2) = |\A_{N_1,N_2}|^2$, where the quantum amplitude is given by $\A_{N_1,N_2} = {}_{12}\IP{N_1,N_2}{1,\beta}^{out}_{12}$. For the input state $\ket{1,\beta}^{in}_{12} = a^\dag_1\,D_2(\beta)\ket{0,0}_{12}$, where 
$D_2(\beta) = e^{\beta\,a^\dag_2 - \beta^*\,a_2}$ is the mode-2 displacement operator, defined such that its action on the vacuum state produces a CS,
$D_2(\beta)\ket{0}_2=\ket{\beta}_2$ \cite{Scully_Zubairy:1997,Agarwal:2013,Ou_Book:2017,Gerry_Knight:2023}. The action of the BS is affected by writing the input creation operators in terms of the output creation operators (and again, dropping $in$ and $out$ subscripts for clarity) producing $\ket{1,\beta}^{out}_{12} = \frac{1}{\sqrt{2}} (a^\dag_1+a^\dag_2)\,D_1(\tfrac{-\beta}{\sqrt{2}})\,D_2(\tfrac{\beta}{\sqrt{2}})\, \ket{0,0}_{12} =\frac{1}{\sqrt{2}} (a^\dag_1+a^\dag_2), \ket{\tfrac{-\beta}{\sqrt{2}},\tfrac{\beta}{\sqrt{2}}}_{12} $, where the transformation of $D_2(\beta)$ by the BS, and the independence of mode-1 and mode-2 operators, creates the (tensor) product of displacement of operators
$D_1(\tfrac{-\beta}{\sqrt{2}})\,D_2(\tfrac{\beta}{\sqrt{2}})$, leading to a product of mode-1/mode-2 coherent states, each with reduced amplitudes $\tfrac{-\beta}{\sqrt{2}}$ and $\tfrac{\beta}{\sqrt{2}}$, respectively. Using the Hermitian conjugate of the boson operator relations that $a_i\ket{N}_i = \sqrt{N}\,\ket{N-1}_i$, namely
${}_{i}\bra{N}\, a^\dag_i = \sqrt{N}\,{}_{i}\bra{N-1}$, it is easy to show that 
\be{P:N1:N2}
P(N_1,N_2) = |{}_{12}\IP{N_1,N_2}{1,\beta}^{out}_{12}|^2
= \frac{e^{-|\beta|^2}}{N_1!\,N_2!\,2^{N_1+N_2}}\,(N_1-N_2)^2,   
\ee
which is a straightforward extension \cite{Ou:1996:Eq:4:note} of a  result first shown by Ou in 1996 \cite{Ou:1996} 
for the input state $\ket{1,N}_{12}$. The salient point here is that for $N_1=N_2$ one has a CNL of zeros along the main diagonal of the joint output probability distribution $P(N_1,N_2)$.

To model imperfect detection efficiency \cite{Loudon:1983,Scully_Zubairy:1997, Knight:2002,eHOM_PRA:2022}
consider the detection of $n$ photons in the output of a single mode, with detector efficiency $0\le\eta\le1$. 
The relevant point is that these $n$ detected photons could have resulted from $N\ge n$ photons impinging on the detector, of which only $n$ were actually registered, due the finite efficiency of the detector. 
The the probability that $n$ photons were detected, which $N-n$ were not, is the Bernoulli factor
$\binom{N}{n}\,\eta^n\,(1-\eta)^{N-n}$, where the binomial coefficient indicates the indistinguishability of which $n$ of the total of the impinging $N$ photons were actually detected. The total probability $P_\eta(n)$ is then a sum over all possible values of $N\ge n$, namely 
$P_\eta(n) = \sum_{N=n}^\infty\binom{N}{n}\,\eta^n\,(1-\eta)^{N-n}\,P_N$ , where 
$P_N=|\mathcal{A}_N|^2$ is the probability to perfectly detect $N$ photons 
\cite{quantum:efficiency:note}. 
%
Applying this to each output mode of the BS, and assuming equal detector efficiencies, the joint probability $P_\eta(n,n)$
to measure $n$ coincidence counts from the output of the BS for the input $\ket{1,\beta}_{12}$ is given by
\bea{P:eta:n:n}
P_\eta(n,n) &=&\sum_{N_1=n}^\infty\, \sum_{N_2=n}^\infty\,\binom{N_1}{n}\,\binom{N_2}{n}\,
\eta^{2\,n}\,(1-\eta)^{N_1+N_2-2n}\,P(N_1,N_2), \no
&=& 2\,\eta^{2\,n}\sum_{N_1>N_2\ge n}^\infty\,\binom{N_1}{n}\,\binom{N_2}{n}\,
\eta^{2\,n}\,(1-\eta)^{N_1+N_2-2n}\,P(N_1,N_2)> 0.
\eea
In the last line of \Eq{P:eta:n:n} we have broken the double sum into two pieces: (i) a single diagonal sum over $N_1=N_2$, which is zero, since $P(N_1,N_1)=0$ by \Eq{P:N1:N2}, and (ii) the remaining off-diagonal double sum, now with $N_1>N_2\ge n$, with a factor of $2$ out front (due to the symmetry of the entire expression under the exchange $N_1\leftrightarrow N_2$). Thus, while the presence of the CNL can be observed, it is no longer exactly zero. In addition, detecting the coincidence output state $\ket{n,n}_{12}$ is proportional to $\eta^{2\,n}\ll 1$ at very high $n$. Thus, the prospects for detecting the presence of the CNL is best for very low photon number (e.g. $n\in\{1,2\}$, and suggests the use of photon number resolving detectors, as suggested in \cite{eHOM_PRA:2022}, which can now detect and number-resolve up to 100 individual photons \cite{eaton_pfister_100_photons:2023}. 

\section{Space-time domain considerations}\label{sec:Rempe:calc}
In the previous sections, we have modeled the FS/FS inputs $\ket{n,m}_{12}$ as idealized  monochromatic, single frequency entities, with (i) photons from both input ports arriving at the BS simultaneously (no time delay between input photons), and similary (ii) photons detected in the output ports simultaneously (no detection time difference). Such effects can be included by considering the photons as arriving in wavepackets, with a finite frequency spread about a central frequency, and incorporating detection delay times.   A general analysis in the frequency domain for an  $N$-photon FS $\ket{N}$, with possible arbitrary $k$ sub-groupings of the photons into time-distinguishable groups containing $N_k$ photons such that $\sum_k N_k = N$, was carried out by Ou \cite{Ou:1996,Ou:2007,Ou_Book:2017}. However, for our purposes, it is sufficient to consider all photons in both input modes, to each be in their own single spatio-temporal mode (and hence indistinguishable). We can include the effects of time delays $\delta\tau$ between the input wavepackets of mode-1 and mode-2, and a difference $\tau$ in the detection time  of the output mode-1 and mode-2 photons, as described in Legero \tit{et al.} \cite{Legero_Rempe:2003}. While this analysis can be carried out in either the frequency or the space-time domain, the authors \cite{Legero_Rempe:2003} advocated that there is less computation involved when using the time domain, as we shall also use here.
Lastly, for simplicity, in this section we will assume unit detection efficiency. The effects of finite detection efficiency can be straightforwardly included afterwards using the Bernoulli trial analysis of the previous section.

The main idea \cite{Legero_Rempe:2003} is that one assumes the Hilbert space of states is spanned by an orthonormal set of spatio-temporal modes $\zeta_k(t)$ such that the positive and negative electric field operators $E^+(t)$ and  $E^-(t)$ (with total electric field $E(t) =E^+(t)+E^-(t)$) are given by
 $E^+(t) = \sum_k \zeta_k\,a_k$ and $E^-(t) = \sum_k \zeta_k\,a^\dag_k$. Then, all that has to be modified from  \Eq{BS:S:out:in:terms:of:in} for a 50:50 BS is
 \be{BS:S:out:in:terms:of:in:eta}
\hspace{-0.5in}
\left[
\begin{array}{c}
E^+_{1,out}  \\
E^+_{2,out}  
\end{array}
\right] = 
\frac{1}{\sqrt{2}}\,
\left[
\begin{array}{c}
E^+_{1,in} - E^+_{2,in}  \\
E^+_{2,in} + E^+_{1,in}  
\end{array}
\right]
=
\frac{1}{\sqrt{2}}\,
\left[
\begin{array}{c}
\zeta_1(t)\,a_{1,in} -\zeta_2(t)\,a_{2,in}\\
\zeta_2(t)\,a_{2,in} + \zeta_1(t)\,a_{1,in} 
\end{array}
\right].
\ee

\subsection{Space-time analysis of the HOM effect: A brief summary of Legero \tit{et al.} \cite{Legero_Rempe:2003}}\label{subsec:HOM:Rempe:Summary}
Consider the two-photon HOM input state $\ket{\Psi_{in}}_{12} = \ket{1,1}_{12} = a^\dag_{1,in}\,a^\dag_{2,in}\,\ket{0,0}_{12}$. The from Glauber's detection theory \cite{Loudon:1983,Loudon:2000},
the joint probability $P_{1_1,1_2}(t_0,\tau)$ to detect an output photon in mode-1 at (arbitrary) time $t_0$, and the output photon in mode-2 at time $t_0+\tau$ is given by 
$P_{1_1,1_2}(t_0,\tau) = |\IP{\Psi_{out}}{\Psi_{out}}|^2$ where 
$\ket{\Psi_{out}}_{12}  = E^+_{2,out}\,E^+_{1,out}\ket{\Psi_{in}}_{12}$ using \Eq{BS:S:out:in:terms:of:in:eta}, which annihilates the two photons at two different times, separated by the time difference $\tau$. Using the boson operator identity for any mode-$i$, 
$a_i\,a^\dag_i \equiv [a_i, a^\dag_i] + a^\dag_i\,a_i = 1 + a^\dag_i\,a_i$ and $a_i\ket{0}_i=0$, and the independence of mode-1 and mode-2, one can push the creation operators in $\ket{\Psi_{in}}_{12}$ to the left of the annihilation operators in electric field operators $E^+_{2,out}\,E^+_{1,out}$ (i.e. arranging the operators in \tit{normal order}), to see that only the commutators $[a_i, a^\dag_i]\ket{00}_{12} = \ket{00}_{12}$ contribute to $\ket{\Psi_{out}}_{12}$ (since in normal order the  annihilation operators will annihilate the vacuum state).
A straightforward calculation produces 
$\IP{\Psi_{out}}{\Psi_{out}} = \half \big(\zeta_1(t_0+\tau)\,\zeta_2(t_0) - \zeta_2(t_0+\tau)\,\zeta_1(t_0)\big)$ for a joint output probability given by Legero \tit{et al.} \cite{Legero_Rempe:2003}
\be{P:joint:HOM}
P^{(HOM)}_{1_1,1_2}(t_0,\tau) = 
\tfrac{1}{4}\,|\zeta_1(t_0+\tau)\,\zeta_2(t_0) - \zeta_2(t_0+\tau)\,\zeta_1(t_0)|^2.
\ee
As pointed out by the authors, if the photons are detected simultaneously 
($\tau=0$), then $P^{(HOM)}_{1_1,1_2}=0$ regardless of the form of the spatio-temporal mode functions.

The above assumes the detection time difference $\tau$ is much smaller than the mutual coherence time 
$\tau_c$ of the incoming photons. If this is not the case, then the interference terms in \Eq{P:joint:HOM} washes out when averaged over an ensemble of different photon pairs and one has 
$P^{(HOM)}_{1_1,1_2}(t_0,\tau)\overset{\tau\gg\tau_c}{\longrightarrow}
\tfrac{1}{4}\,\big(P_1(t_0+\tau)\,P_2(t_0) + P_2(t_0+\tau)\,P_1(t_0)\big)$
where $P_i(t) = {}_{12}\bra{\Psi_{in}} E^-(t)\,E^+(t) \ket{\Psi_{in}}$ is the probability (proportional to the intensity) to measure the photon in output port-$i$.

To study quantum beating between the two-photons in the HOM effect, Legero \tit{et al.} \cite{Legero_Rempe:2003} used the normalized spatio-temporal Gaussian mode functions 
($\int_{-\infty}^{\infty} |\zeta_i(t)|^2\,dt = 1$)
\bsub
\bea{Rempe:24}
\zeta_1 &=& \sqrt[4]{2/\pi}\; e^{-(t-\delta\tau/2)^2 - i\,(\omega - \Delta\omega/2)t}, \\
\zeta_2 &=& \sqrt[4]{2/\pi}\; e^{-(t+\delta\tau/2)^2 - i\,(\omega + \Delta\omega/2)t},
\eea
\esub
where $\omega = \half\,(\om_1+\om_2)$ and $\Delta\om = (\om_1-\om_2)$ are the average and difference between the two mode-1 and mode-2 frequencies $\om_1$ and $\om_2$, respectively.
Here, $\delta\tau$ is the total time delay between the center frequencies of the mode-1 and mode-2 input wavepackets.
The total probability is obtained by integrating over all possible values of the arbitrary detection time $t_0$ of the first photon and is given by \cite{Legero_Rempe:2003}
\be{Rempe:26}
\hspace{-0.2in}
P^{(HOM)}_{1_1,1_2}(\tau, \delta\tau, \Delta\om) = \int \, dt_0\, P^{(HOM)}_{1_1,1_2}(t_0,\tau, \delta\tau, \Delta\om) = \frac{\cosh(2\tau\,\delta\tau)-\cos(\Delta\om\,\tau)}{2\,\sqrt{\pi}}\; 
e^{-(\tau^2+\delta\tau^2)}.
\ee 
Other realistic experimental effects can be incorporated such as (i) inhomogeneous broadening of the frequency difference (using a normalized Gaussian distribution 
$f_{\delta\om}(\Delta\om) = \tfrac{1}{\delta\om\,\sqrt{\pi}}\,e^{-(\Delta\om/\delta\om)^2}$), and (ii) integrating over the detection time difference $\tau$ to obtain 
\be{Rempe:29}
\hspace{-0.2in}
P^{(HOM)}_{\trm{total}; 1_1,1_2}(\delta\tau, \delta\om) = 
\int\int \, d\tau\,d\Delta\om\, f_{\delta\om}(\Delta\om)\, P^{(HOM)}_{1_1,1_2}(\tau, \delta\tau, \Delta\om) 
= \half - \frac{e^{-\delta\tau^2}}{\sqrt{4+\delta\om^2}}\overset{\delta\tau\gg\sigma_t}{\longrightarrow} \half.
\ee 
\Eq{Rempe:29} shows that the HOM effect vanishes if the two wavepackets are delayed in time by greater than the width of their wavepackets $\sigma_t$, since they are then distinguishable photons.

\subsection{Space-time analysis of the Fock state/coherent state input state $\ket{1,\beta}_{12}$}\label{subsec:HOM:Rempe:Summary}
In this section we once again analyze the FS/CS input state $\ket{1,\beta}_{12}$, and the possibility to observe the eHOM effect on the output coincidence state $\ket{1,1}_{12}$, 
but now from a space-time analysis, as in the previous section.
We take as our spatio-temporal mode functions
\be{PMA:zeta:1:zeta:2}
\zeta_1(t) = \sqrt[4]{\frac{2}{\pi\,\tau^2_c}}\; e^{-i\,\om_1\,t - t^2/\tau^2_c}, \qquad
\zeta_2(t) = \sqrt{\mathcal{F}} \,e^{-i\,\om_2\,t}\,e^{i\,\theta}.
\ee
In the above, we have modeled the CW laser as a single frequency exponential of frequency $\om_2$ with phase $\theta$,  following Loudon \cite{Loudon:2000}. 
Here,  $\mathcal{F}$ is the laser flux such that in frequency space
$f(\om) =  \mathcal{F}\,\delta(\om-\om_2)$ and $\EV{a^\dag_2\,a_2}{12}  = \int d\om\, f(\om) = \mathcal{F} 
= \int dt\, f(t)$.
For the single mode-1 input wavepacket,  we have taken a normalized Gaussian of FWHM\; 
$\tau_c = 1/\Delta\om$ where $\Delta\om$ is the frequency bandwidth of the mode-1 wavepacket with center frequency $\om_1$. 

Since the mode-2 input is an idealized  CW laser, there is no time delay $\delta\tau$ between mode-1 and mode-2 - i.e. the monochromatic laser is technically an infinite width pulse in time. If we instead wanted to consider a laser pulse, we could use a similar normalized Gaussian mode function for mode-2 as in \Eq{PMA:zeta:1:zeta:2} with temporal width $\tau_{2,c}$, and consider 
$\tau_{2,c}\gg \tau_{1,c}$ so that the mode-1 wavepacket is always ``under" the mode-2 laser ``wavepacket." To keep things simple, we  instead use the mode functions in \Eq{PMA:zeta:1:zeta:2}, effectively taking
$\tau_{2,c}\to\infty$ and writing $\tau_{1,c}\to\tau_{c}$. The analysis then proceeds in a similar fashion to the HOM case considered previously, 
but now for the FS/CS input state
$\ket{\Psi_{in}}_{12} =\ket{1,\beta}_{12} = a^\dag_{1,in}\,\ket{0,\beta}_{12}$.

We then form (upon dropping the $in$ subscripts on the operators), 
\bea{Psi:in:1:beta}
\ket{\Psi_{out}}_{12} &=& E_{2,out}^{+}(t_0+\tau)\, E_{1,out}^{+}(t_0)\,a^\dag_{1,in}\,\ket{0,\beta}_{12}\no
&=& \half
\Big[
\zeta_1(t_0+\tau)\zeta_1(t_0) \, a^2_1 -
\zeta_2(t_0+\tau)\zeta_2(t_0)\, a^2_2  \no
&-& 
\big(\zeta_1(t_0+\tau)\zeta_2(t_0) - \zeta_2(t_0+\tau)\zeta_1(t_0)\big)\,a_1\,a_2 
\Big]\, 
a^\dag_1\,\ket{0,\beta}_{12}, \no
&=&
-\half\, 
\Big[
\zeta_2(t_0+\tau)\zeta_2(t_0)\, \beta^2 a^\dag_1 + 
(\zeta_1(t_0+\tau)\zeta_2(t_0) - \zeta_2(t_0+\tau)\zeta_1(t_0))\,\beta 
\Big]\,\ket{0,\beta}_{12},\qquad
\eea
where the term linear in $\beta$ is the same interference amplitude that appeared in the HOM calculation in \Sec{subsec:HOM:Rempe:Summary}. In going from the second to the third equality, we have used the fact that $a^2_1\,a^\dag_1 \equiv [a^2_1, a^\dag_1] + a^\dag_1\,a^2_1 = 2\, a_1 + a^\dag_1\,a^2_1$, with both terms acting to the right, annihilating the mode-1 vacuum $\ket{0}_1$.
When forming $P_{1_1,1_2}(t_0,\tau) = |\IP{\Psi_{out}}{\Psi_{out}}|^2$ only ``like" powers of $(a^\dag_1)^k$ for $k\in\{0,1\}$ contributed to the final result (again using $a_1\,a^\dag_1 = 1 +a^\dag_1\,a_1)$, since all other terms involve lone operators $a_1$ or $a^\dag_1$ which annihilate the mode-1 vacuum when acting to the right or left, respectively. Thus, we obtain
\bsub
\bea{Psi:in:1:beta:result}
 P^{\ket{1,\beta}_{12}}_{1_1,1_2}(t_0,\tau) &=& 
 \frac{1}{4}
\left(
 P^2_2(t_0)\,\bar{n}_2^2 +
 |\zeta_1(t_0+\tau)\zeta_2(t_0) - \zeta_2(t_0+\tau)\zeta_1(t_0)|^2\,\bar{n}_2
 \right),  \label{Psi:in:1:beta:result:1} \label{Psi:in:1:beta:result:1} \\
 %
 =\frac{1}{4}\,F\,\bar{n}_2^2 
 &+& 
 2\,F\,\bar{n}_2\,\sqrt{\frac{2}{\pi\,\tau^2_c}}\, e^{-\tau^2/2\,\tau^2_c}\, e^{-2\,(t_0+\tau/2)^2/2\,\tau^2_c}\,
 \Big(
 \cosh(\tau(2\,t_0+\tau)/\tau^2_c) - \cos(\Delta\om\,\tau)
 \Big),\qquad\;\; \label{Psi:in:1:beta:result:2}
\eea
\esub
where $\Delta\om\defn \om_1-\om_2$ and 
$P_2(t_0) = P_2(t_0+\tau)$ is the probability to detect a photon from the mode-2 input CS 
$\ket{\beta}_{2}$, with mean number of photons $\bar{n}_2 = |\beta|^2$.
The first non-interfering, DC term in \Eq{Psi:in:1:beta:result:1} arises from the squared amplitude for the process in which both of the output photons in the two-photon coincidence detection come from the CS. 
The second, interference term, is the probability that the output coincidence pair consists of one photon from each mode.
The origin of this term is somewhat obscured by the use of a CS, which is an eigenstate of the mode-2 annihilation operator, 
$a_2\ket{\beta}_2 = \beta\,\ket{\beta}_2$. A calculation similar to \Eq{Psi:in:1:beta}, but now with input state 
$\ket{n,m}_{12}$, produces three amplitudes 
$\ket{\Psi_{out}}_{12}^{\ket{n,m}_{12}} = \mathcal{A}_1 + \mathcal{A}_2 + \mathcal{A}_3$
with 
(i) $\mathcal{A}_1 = \tfrac{1}{4}\,\big(\zeta_1(t_0+\tau)\,\zeta_1(t_0)\,\sqrt{n(n-1)}\big)$, the DC amplitude for both of the output coincidence photons to come from mode-1, 
(ii) $\mathcal{A}_2 =  \tfrac{1}{4}\,\big(\zeta_2(t_0+\tau)\,\zeta_2(t_0)\,\sqrt{m(m-1)}\big)$, the DC amplitude for both of the output coincidence photons to come from mode-2, and finally
(iii) $\mathcal{A}_3=-\tfrac{1}{4}\,\big(\zeta_1(t_0~+~\tau)\,\zeta_2(t_0) - \zeta_2(t_0~+~\tau)\,\zeta_1(t_0) \big)\, \sqrt{n\,m}$, the HOM interference amplitude for the  output coincidence pair to consist of one photon from mode-1 and the other photon from mode-2.

Coming back to the calculation yielding \Eq{Psi:in:1:beta:result:1} with input state $\ket{1,\beta}_{12}$,
note that the first DC term in \Eq{Psi:in:1:beta:result} is multiplied by $(\bar{n}_2)^2$, since this term arises from both of the  detected photons coming from the CS, 
while the second, eHOM interference term is multiplied by only $\bar{n}_2$, since then only one photon comes from the CS. Thus, in order to observe the eHOM effect on the coincidence output state
$\ket{1,1}_{12}$, one would have to measure   the first term $\tfrac{1}{4}P^2_2(t_0)\,\bar{n}_2^2$ 
from the input mode-2 CS alone (from direct photon counting measurements), and subtract this off from $P^{\ket{1,\beta}_{12}}_{1_1,1_2}(t_0,\tau)$ in \Eq{Psi:in:1:beta:result} in order to observe the HOM-like interference term proportional to $\bar{n}_2$. The total probability for detecting one photon in each of output ports 1 and 2 with a time difference of $\tau$
is given by integrating the HOM interference term in 
\Eq{Psi:in:1:beta:result:2} over all (arbitrary) values of $t_0$, which yields
\be{Psi:in:1:beta:results:int:over:t0}
P^{\ket{1,\beta}_{12}}_{1_1,1_2}(\tau) = \frac{1}{4}\,F\,\bar{n}_2^2  + 
\half\, F\,\bar{n}_2 
\left(
1 - e^{-\tau^2/2\,\tau^2_c}\,\cos(\Delta\om \,\tau)
\right).
\ee
Again, for $\tau=0$ the HOM-like interference term in \Eq{Psi:in:1:beta:results:int:over:t0} is zero, regardless of the frequency difference $\Delta\om$.
In the other extreme, the second term in \Eq{Psi:in:1:beta:results:int:over:t0} reduces to $\half\,F\,\bar{n}_2$ if  the second term in the parentheses either
%
(i) vanishes when $\tau\gg\tau_c$, even if $\Delta\om=0$ (identical frequencies),   or  
(ii) washes out if $\tau\gg 1/\Delta\om$.
Thus, the prospects for observing the eHOM effect on the coincidence output state $\ket{1,1}_{12}$ for the FS/CS input state $\ket{1,\beta}_{12}$ is optimal for indistinguishable photons detected simultaneously.
\subsection{Higher Order Multi-photon Counting}\label{sec:Kelly:Kleiner}
In this last section we provide an indication of how higher order multi-photon counting is computed using the 
classic quantum photon counting formula developed by Kelly and Kleiner \cite{Kelly_Kleiner:1964,Loudon:1983,Ou:1996}.
This formula would be necessary for example, for measuring the output coincident state $\ket{N,N}_{12}$ for the FS/CS input state $\ket{1,\beta}_{12}$, considered in the last section, for $N>1$.
For a two-port device one, has for the probability to detect $N_1$ photons in mode-1, and $N_2$ photons in mode-2, with detector efficiencies $\eta_1$ and $\eta_2$, respectively\cite{Ou:1996}, is given by 
%
\be{Ou:3}
\hspace{-0.65in}
P_{\eta_1,\eta_2}(N_1,N_2) = 
\Tr\left[\rho_{12}\,
:
\tfrac{\left(\eta_1\,a^\dag_{1,out}\,a_{1,out}\right)^{N_1}\,e^{-(\eta_1\,a^\dag_{1,out}\,a_{1,out})}}{N_1!}\,
\tfrac{\left(\eta_2\,a_{2,out}\,a^\dag_{2,out}\right)^{N_2}\,e^{-(\eta_2\,a^\dag_{2,out}\,a_{2,out})}}{N_2!}
:
\right],
\ee
where $\rho_{12}$ is a general bipartite  density matrix (we've been considering the pure cases when
 $\rho_{12}~=~\ket{\Psi_{in}}_{12}\bra{\Psi_{in}}$). In \Eq{Ou:3} the  double colons 
 $:O(a, a^\dag):$ are the conventional notation for putting the operator  $O(a, a^\dag)$ in 
 \tit{normal order}, defined by simply reordering any operator expression by placing all the creation operators to the left, and all annihilation operators to the right, \tit{without} regard for the boson commutation relations. As an example, whereas $(a^\dag\,a)^2 \equiv a^\dag (a\,a^\dag)\,a = 
 a^\dag ([a, a^\dag] + a^\dag\,a)\,a = (a^\dag)^2\,a^2 + a^\dag\,a$, 
 for normal ordering we simply have
 $:(a^\dag\,a)^2:~\defn~(a^\dag)^2\,a^2$ directly, by reshuffling (normal ordering) the operators.
 The process of normal ordering is appropriate for physical detectors, since in expressions such as
 ${}_{in}\EV{(a^\dag)^2\,a^2}{in}$, the annihilation operators (acting to the right) only remove photons from the input state 
 $a^2\ket{\Psi_{in}}$ (as opposed to creating photons in the detector), and the 
 term $\bra{\Psi_{in}}(a^\dag)^2$ is the corresponding Hermitian conjugate of the previous term, which (acting to the left, only) removes photons from $\bra{\Psi_{in}}$.
 
 For a 50:50 BS
 (i) the output creation and annihilation operators in  \Eq{Ou:3} 
 are replaced by their expressions in terms of the input operators, as in \Eq{BS:S:out:in:terms:of:in:eta}, (with or without the spatio-temporal mode functions), 
 (ii) the powers and exponentials of operators are expanded in terms of (binomial) series, and finally 
 (iii) all operators in this complicated expression are normally ordered. 
 \subsubsection{A single mode example}
 To illustrate the above procedure, consider (for simplicity) just the single mode-1, and no BS, i.e. let $a_{1,out}\to a$ and $\rho_{12}\to\rho$ and $\eta_1\to\eta$ (this is just a detector placed in front of the input mode-1 state). Taking the trace in \Eq{Ou:3} over FS $\ket{N}$, we have schematically  
 $\Tr[\rho : O(a^\dag\,a):]$ where $O(a^\dag\,a)$ is a function of the number operator and hence diagonal in the number basis. 
 Therefore, 
 $\Tr[\rho : O(a^\dag\,a):] = \sum_N P_N \left(:O(a^\dag\,a):\right)_{N}$ where 
 $P_N \defn \bra{N} \rho\ket{N}$ and $\left(:O(a^\dag\,a):\right)_{N}\defn$ $\bra{N}:O(a^\dag\,a):\ket{N}$.
 Thus, the single mode version of \Eq{Ou:3} is \cite{Loudon:1983}
 \bea{Loudon:1983:p240}
 P_\eta(n) &=& \sum_N\, P_N \bra{N}:\, \frac{(\eta\,a^\dag\,a)^n}{n!}\,\sum_{k=0}^\infty \frac{(-\eta\,a^\dag\,a)^k}{k!}\,:\ket{N}, \no
 &=& \sum_N \,P_N \, \frac{\eta^n}{n!}\,\sum_{k=0}^\infty \frac{(-\eta)^k}{k!}\,
  \bra{N} (a^\dag)^{n+k}\, a^{n+k} \ket{N}, \no
  &=& \sum_N P_N \, \frac{\eta^n}{n!}\,\sum_{k=0}^\infty \frac{(-\eta)^k}{k!}\,
  \frac{N!}{(N-(n+k))!}, \no
  &\equiv& \sum_{N=n}^\infty P_N \, \eta^N\,\frac{N!}{n!(N-n)!}\,
  \sum_{k=0}^{N-n} \frac{(N-n)!}{k!(N-(n+k))!}\,(1)^{(N-(n-k))}(-\eta)^k  \no
  &=& \sum_{N=n}^\infty P_N \binom{N}{n} \eta^n\,(1-\eta)^{N-n},
 \eea
 where in the second line we have normally ordered the operators before taking the expectation value with respect to the FS $\ket{N}$.
 In the third line we have used $a^{n+k} \ket{N} = \sqrt{\tfrac{N!}{(N-(n+k))!}}\,\ket{N-(n+k)}$, and in the penultimate line we have multiplied and divided by $(N-n)!$,  and have used
 the binomial expansion of $(1-\eta)^{N-n}$ to form the last line.
 \Eq{Loudon:1983:p240} is the single mode version of \Eq{P:eta:n:n}.
 A straightforward calculation reveals that the mean number $\bar{n}$ of photons registered by the detector is given by $\bar{n}\defn \sum_n n\,P_\eta(n)  =  \eta\, \bar{N}$, 
 where $\bar{N}\defn \sum_N N\, P_N$ is the mean number of photons from the input source (as measured by an idealized, unit efficiency detector). 
 
When a BS is involved the $out$ operators in \Eq{Ou:3} must be replaced by their expressions in terms of the $in$ operators per \Eq{BS:S:out:in:terms:of:in:eta} (with or without the mode functions). 
With $\eta_1=\eta_2\defn\eta$, the normal ordered expression of the operators in \Eq{Ou:3} is given by
\be{Ou3:NO}
\frac{\eta^{N_1+N_2}}{N_1!\,N_2!}\,\sum_{k=0}^\infty \sum_{\l=0}^\infty\,\frac{(-\eta)^{k+\l}}{k!\,\l !}\,
(a_{1,out}^\dag)^{N_1+k}\,(a_{2,out}^\dag)^{N_2+\l}\,
(a_{1,out})^{N_1+k}\,(a_{2,out})^{N_2+\l}\,
\ee
leading to the joint output probability 
\bsub
\bea{P:eta:N1:N2:general}
P_\eta(N_1,N_2) &=&  \frac{\eta^{N_1+N_2}}{N_1!\,N_2!}\,\sum_{k=0}^\infty \sum_{\l=0}^\infty\,\frac{(-\eta)^{k+\l}}{k!\,\l !}\, {}_{12}\IP{\Psi^{k\l}_{out}}{\Psi^{k\l}_{out}}_{12}, \label{P:eta:N1:N2:general:1} \\
\ket{\Psi^{k\l}_{out}}_{12} &\defn&(a_{1,out})^{N_1+k}\,(a_{2,out})^{N_2+\l}\,\ket{\Psi_{in}}_{12}, \no
&=& \frac{1}{\sqrt{2^{N_1+N_2+k+\l}}}\,
\sum_{r=0}^{N_1+k} \sum_{s=0}^{N_2+\l} \,\binom{N_1+k}{r}\,\binom{N_2+\l}{r} (-1)^{N_1+k-r}\, \no
&\times&
\zeta^r_1(t_0)\,\zeta^s_1(t_0+\tau)\,
\zeta^{N_1+k-r}_2(t_0)\,\zeta^{N_2+\l-s}_2(t_0+\tau)\,
a_1^{r+s}\,a_2^{N_1+N_2+k+\l -(r+s)}\,
\ket{\Psi_{in}}_{12}, \quad\qquad \label{P:eta:N1:N2:general:2}
\eea
\esub
where we have dropped the $in$ subscript on the annihilation operators after employing \Eq{BS:S:out:in:terms:of:in:eta}.

\subsubsection{Balanced beamsplitter with $\ket{1,\beta}_{12}$ input state}
If we now consider the input state $\ket{\Psi_{in}}_{12} = \ket{1,\beta}_{12}$ with one photon in mode 1, we see that there are only two sets of  contributions in the amplitude sums $a_1^{r+s}\,\ket{1}_1$
 in \Eq{P:eta:N1:N2:general:2}, namely
$(r,s)_1= (0,0)$ and $(r,s)_2\in \{(0,1), (1,0)\}$. The first term $(r,s)_1= (0,0)$ yields a DC contribution 
$P^{(1)}_\eta(N_1,N_2)$ to the joint output probability 
$P_\eta\defn P^{(1)}_\eta+P^{(2)}_\eta$ given by
\be{P:1:eta:N1:N2}
P^{(1)}_\eta(N_1,N_2) = \frac{p^{N_1}\,e^{-p}}{N_1!}\, \frac{p^{N_2}\,e^{-p}}{N_2!}\, P^2_2(t_0)\,\bar{n}_2,
\qquad p \defn \half\,\eta\,\bar{n}_2\,P_2(t_0),
\ee
 where we have used $P_2(t_0) = |\zeta_2(t_0)|^2 = P_2(t_0+\tau)$ for the CS $\ket{\beta}_2$, using 
 \Eq{PMA:zeta:1:zeta:2}.
 For this contribution  all the detected photons come from the CS 
 since $(r,s)_1= (0,0)$ indicates that no mode-1 photons are annihilated by the detector, and the 
 resulting state in the amplitude \Eq{P:eta:N1:N2:general:2} is proportional to $\ket{1,\beta}_{12}$,

 The second set of amplitude contributions in  \Eq{P:eta:N1:N2:general:2} arises from the pair
 $(r,s)_2\in \{(0,1), (1,0)\}$, and  yields a state proportional to $\ket{0,\beta}_{12}$. Here, the mode-1 photon is annihilated along with mode-2 photons, allowing for the possibility of interference. 
 The two amplitude contributions have a relative minus sign due to the factor 
 $(-1)^r = \{1,-1\}$ for $r=\pm 1$ in \Eq{P:eta:N1:N2:general:2} for the two terms. The total joint probability is then given by
 \bea{P:eta:N1:N2:tau}
 P_\eta(N_1,N_2) &=& \frac{\eta^2\,\bar{n}_2}{4 N_1 N_2} \frac{p^{N_1-1}}{(N_1-1)!}\,\frac{p^{N_2-1}}{(N_2-1)!}\,
 \sum_{k=0}^\infty\, \sum_{\l=0}^\infty\,\frac{(-p)^k}{k!}\,\frac{(-p)^{\l}}{\l!}\, \no
 &\times&
 \left[
 P_2(t_0)\,P_2(t_0+\tau)\,\bar{n}_2 + \left|(N_2+\l)\,\zeta_1(t_0+\tau)\,\zeta_2(t_0) - (N_1+k)\,\zeta_1(t_0)\,\zeta_2(t_0+\tau)\right|^2
 \right], \qquad \\
 &\defn& P^{(1)}_\eta(N_1,N_2)+P^{(2)}_\eta(N_1,N_2). \nonumber
 \eea
 The interference term appears as the second term in the large square parentheses in \Eq{P:eta:N1:N2:tau},
 which is readily apparent when we set 
 $N_1=N_2=N$ to examine the diagonal of the joint output probability.
 The second term in the square brackets becomes
 $\left| N\mathcal{A} + \l\,\zeta_1(t_0+\tau)\,\zeta_2(t_0) - k\,\zeta_1(t_0)\,\zeta_2(t_0+\tau)\right|^2$
 with $\mathcal{A}  = \zeta_1(t_0+\tau)\,\zeta_2(t_0)-\zeta_1(t_0)\,\zeta_2(t_0+\tau)$ the HOM interference amplitude from the previous sections.

 The somewhat involved expression for $P_\eta(N_1,N_2) $ in \Eq{P:eta:N1:N2:tau} arises from the fact that we are asking for the interference from an input state $\ket{1,\beta}_{12}$ with only a single photon in mode-$1$, yet we are observing $N_1$ photons in the output of mode-$1$. Thus, $N_1-1$ photons in the observed output mode-$1$ must be coming from the input CS in mode-$2$. The interference term  in \Eq{P:eta:N1:N2:tau} describes all the possible ways that the single mode-$1$ input photon can interfere with each of the $N_1-1$ CS mode-$2$ input photons.
 
To make connection to the idealized eHOM effect in \Sec{sec:Review:HOMisOdd} (and in \cite{eHOM_PRA:2022}), we set $\tau=0$, which in all practical purposes is experimentally realizable to within a full width half maximum set by the ``jitter time" $\tau_{\trm{jitter}}$ of the detector, i.e. 
$\tau\ge \tau_{\trm{jitter}}$, 
where $\tau_{\trm{jitter}}$ can be as low as $\sim 15-25 \trm{ps}$  for some detectors. 
Since $\mathcal{A}\overset{\tau\to0}{=}0$ regardless of the form of the mode functions $\zeta_1$ and $\zeta_2$, the interference term in the square brackets above reduces to $P_1(t_0)\,P_2(t_0)\,(\l-k)^2$.
 Using $\sum_{k=0}^\infty k \tfrac{(-p)^k}{k!}=(-p)\,e^{-p}$ and  
 $\sum_{k=0}^\infty k^2 \tfrac{(-p)^k}{k!}=p(p-1)\,e^{-p}$ (and similarly for the sums over $\l$),
 the joint output probability in \Eq{P:eta:N1:N2:tau} for $N_1=N_2=N$ reduces to
 \bsub
 \bea{P:eta:N:N:tau:0}
 P_\eta(N,N) &\overset{\tau\to0}{=}& 
 \frac{\eta^2\,\bar{n}_2}{4 N^2} \left(\frac{p^{N-1}}{(N-1)!}\right)^2\,
 e^{-2\,p}\,P^2_2(t_0)\,\bar{n}_2\,[1-\eta\,P_1(t_0)], \label{P:eta:N:N:tau:0:line:1} \\
&{}& \overset{\eta,\, P_1\to 1}{\longrightarrow} 0. \label{P:eta:N:N:tau:0:line:2}
 \eea
 \esub
 \Eq{P:eta:N:N:tau:0:line:2} indicates that in the idealized limit of 
 (i) perfect detection efficiency $\eta\to 1$,  (ii) no time delay in the coincidence detection $\tau\to 0$, and  (iii) ignoring the spatio-temporal mode function $\zeta_1\to 1$ ($P_1 = |\zeta_1|^2\to 1$) of the input single mode-1 photon $\ket{1}_1$ (all approximations implicitly assumed  in \cite{eHOM_PRA:2022}, and in \Sec{sec:Review:HOMisOdd}), there exist a central nodal line (CNL) of zeros along the diagonal of the joint output probability distribution $P(N,N) =0$ for the FS/CS input state $\ket{1,\beta}_{12}$.
 
 The analysis for  arbitrary input state $\ket{\Psi_{in}}_{12}$, proceeds similarly using
 \Eq{P:eta:N1:N2:general:1} and \Eq{P:eta:N1:N2:general:2}. However, for more than one photon in mode-1, care must be taken to account for all the amplitude contributions from $a_1^{r+s}$ acting on $\ket{\Psi_{in}}_{12}$, before the sums over $k$ and $\l$ are performed.

 
 \section{Summary and Conclusion}\label{sec:Conclusion}
 The two photon HOM effect with input state $\ket{11}_{12}$ to a lossless 50:50 BS with detection on the output coincidence state $\ket{11}_{12}$ reveals that the total quantum amplitude contains two terms, of equal magnitude but of opposite sign, hence summing to zero. By default of the choice of the input state, the two amplitude contribution diagrams in question are automatically ``mirror-images" of each other, namely swapping the number of photon transmitted/reflected by the mode-1 photon, with the number reflected/transmitted. 
 
 What the eHOM effect unveils is that for higher order odd-odd FS/FS inputs 
 $\ket{n,m}_{12}$, the total amplitude for the output coincident state 
 $\ket{\tfrac{n+m}{2}, \tfrac{n+m}{2}}_{12}$ consists of a sum of pairs of mirror-image amplitude contribution diagrams, again swapping the number of photons transmitted/reflected by the mode-1 photon, with the number reflected/transmitted, which have equal magnitude, but opposite sign. Thus, the total amplitude cancels pairwise on multiple  mirror-image amplitude contributions. In a sense, this is a series of  HOM-like effects on each of the $\half(n+1)$ mirror-image pair amplitude contributions.
 
 For even-even FS/FS inputs $\ket{n,m}_{12}$, the total amplitude for the output coincident state 
 contains an odd number of terms consisting of $\half\,n$ mirror-image pairs, and a lone un-paired ``middle" term. However, since this time $n$ is even, the mirror-image pair amplitude contributions are equal in magnitude with the \tit{same} sign, and so constructively (vs destructively) interfere. Regardless, the un-paired middle term is non-zero, so the net result is that for a 50:50 BS, there will always be a non-zero constructive interference, when $n$ is even.
 
 The implication of this result, is that for any odd-parity state (consisting of only odd number of photons) entering the mode-1 input port, then regardless of the state entering the mode-2 input port, be it pure or mixed, there will always be a central nodal line (CNL) of zeros (destructive interference) along the the diagonal of the joint output probability distribution. This the statement of eHOM effect \cite{eHOM_PRA:2022}.
 

In this work we have explained diagrammatically how the eHOM effect can be seen a multi-photon generalization of the standard two-photon HOM effect through successive pairwise cancellation of mirror image scattering amplitudes that occur in the former process.
 We considered the prospects for realization of the eHOM effect on the $\ket{1,1}_{12}$ output coincidence state for the experimentally accessible FS/CS input state 
 $\ket{1,\beta}_{12}$. We first explored the modification of the effect due to imperfect detection efficiency, and then explored the modifications due to frequency difference of the two input states, and a time difference between the output single photon detections. Lastly, we used the Kelly-Kleiner photon counting formula to examine the effects produced by the mode function of the  single input photon, as well as a possible time delay between the detected output coincident photons (of arbitrary $N$). We showed that in the appropriate idealized limit, these more experimentally realistic results reduced to the CNL of zeros (complete destructive interference) in the joint output probability considered in \cite{eHOM_PRA:2022}.
 
 The lesson of the HOM and eHOM effects highlights the fundamental importance of the discreetness of photonic quanta (which can be observed by photon counting) in quantum interference effects, 
 and the power of the non-classicality of quantum states, for even a single photon to have a measurable effect on a macroscopic classical-like state (the CS ``laser" $\ket{\beta}$).
 
\begin{acknowledgments}
Any opinions, findings and conclusions or recommendations expressed in this material are those of the author(s) and do not necessarily reflect the views of their home institutions.
\end{acknowledgments}


\begin{thebibliography}{10}
\bibitem{quantum:efficiency:note}
It is worth noting that the \textit{quantum efficiency} of the detector is
  given by (see Section 6.8 of Loudon \cite{Loudon:1983}) $\xi =
  \eta\,c\,\hbar\omega\,T/V = \eta\,\hbar\omega/\sigma$, where $\eta$ is the
  detector efficiency, $T$ is the integration time of the detector of photons
  of frequency $\omega$, and $V$ is a quantization volume. Here $\sigma =
  V/(T\,c)$ is the effective cross section for the detection process of the
  photons. The use of $\xi$ over $\eta$ allows one to incorporate the effect of
  very long integration times $T$. The net effect is that in the expression for
  $P_\eta(n)$ one replaces $\eta\to 1- e^{-\xi}$ so that $P_\xi(n; T) =
  \sum_{N=n}^\infty\binom{N}{n}\,(1- e^{-\xi})^n\,e^{-(N-n)\,\xi}\,P(N)$, which
  reduces to the expression given in the text when $\xi\ll 1$ and one can
  approximate $1- e^{-\xi}\approx \xi$ and $e^{-(N-n)\,\xi}\approx 1-(N-n)\,\xi
  \approx (1-\xi)^{N-n}$, with $\xi\to\eta$. In the limit of very large $\xi$
  one has $P_\xi(n; T)\overset{\xi\to\infty}{\longrightarrow}
  P(N)\,\delta_{N,n}$. For most detectors in current use with $\eta<1$, the
  expression in the text is adequate and appropriate.
\end{thebibliography}
\providecommand{\noopsort}[1]{}\providecommand{\singleletter}[1]{#1}%

\end{document}